\documentclass[11pt]{article}
\usepackage{amsfonts,amssymb,amsmath}

\usepackage[english]{babel}
\usepackage{color}
\newtheorem{theorem}{Theorem}
\newtheorem{definition}{Definition}

\newtheorem{proof}{Proof}

\newtheorem{proposition}{Proposition}
\newtheorem{lemma}{Lemma}

\newcommand{\beq}{\begin{eqnarray}}
\newcommand{\eeq}{\end{eqnarray}}

\newcommand{\beqt}{\begin{eqnarray*}}
\newcommand{\eeqt}{\end{eqnarray*}}

\newcommand{\be}{\begin{equation}}
\newcommand{\ee}{\end{equation}}

\newcommand{\bl}{\begin{lemma}}
\newcommand{\el}{\end{lemma}}

\newcommand{\bt}{\begin{theorem}}
\newcommand{\et}{\end{theorem}}

\newcommand{\bd}{\begin{definition}}
\newcommand{\ed}{\end{definition}}

\newcommand{\bp}{\begin{proposition}}
\newcommand{\ep}{\end{proposition}}

\newcommand{\bpr}{\begin{proof}}
\newcommand{\epr}{\end{proof}}

\newcommand{\bi}{\begin{itemize}}
\newcommand{\ei}{\end{itemize}}

\newcommand{\ben}{\begin{enumerate}}
\newcommand{\een}{\end{enumerate}}

\newcommand{\Z}{\mathbb Z}

\newcommand{\N}{\mathbb N}

\newcommand{\E}{\mathbb E}

\newcommand{\s}{\ensuremath{\mathcal{S}}}

\newcommand{\om}{\ensuremath{\omega}}
\newcommand{\Om}{\ensuremath{\Omega}}

\newcommand{\La}{\ensuremath{\Lambda}}
\newcommand{\si}{\ensuremath{\sigma}}

\begin{document}

\title{{\bf Dyson models under renormalization and in weak fields}}
 
\author{R. Bissacot \footnote{Institute for Mathematics and Statistics -- IME  USP-- Universidade de S\~ao Paulo, Brazil, 
\newline
email: rodrigo.bissacot@gmail.com}, \\
%\bblue
 E.O. Endo 
 %\ew 
 \footnote{Johann Bernoulli Institute, University of Groningen, Nijenborgh 9, 9747AG, Groningen, Netherlands, and Institute for Mathematics and Statistics -- IME USP -- Universidade de S\~ao Paulo, Brazil, 
\newline
email: eric@ime.usp.br},\\
 A.C.D.  van Enter \footnote{Johann Bernoulli Institute, University of Groningen, Nijenborgh 9, 9747AG,Groningen, Netherlands,
 \newline
 email: aenter@phys.rug.nl},\\ 
  B. Kimura \footnote{Delft Institute for Applied Mathematics, Technical University Delft, Mekelweg 4, 2628CD, Delft,  Netherlands,
  email bruno.hfkimura@gmail.com}, \\
  A. Le Ny \footnote{LAMA UMR CNRS 8050, UPEC, Universit\'e Paris-Est,  94010 Cr\'eteil, France,
  \newline
 email:  arnaud.le-ny@u-pec.fr},\\
 %\footnote{E-mail: aenter@phys.rug.nl,  arnaud.le-ny@u-pec.fr}
W.M. Ruszel \footnote{Delft Institute for Applied Mathematics, Technical University Delft, Mekelweg 4, 2628CD, Netherlands, 
\newline
email: W.M.Ruszel@tudelft.nl}
}

\maketitle

\begin{center}
{\bf Abstract:} 
%{\bf Preliminary Draft}}
\end{center}
%We study the decimation to a sublattice of half the sites of 
We consider one-dimensional long-range spin models (usually called {\em Dyson models}), consisting of 
Ising ferromagnets with a slowly decaying long-range pair potentials of the 
form $\frac{1}{| i-j |^\alpha}$,  mainly focusing  on the range of slow decays $1 < \alpha \leq 2$. 
We describe two recent results, one about renormalization and one about the effect of  external fields at low temperature.\\
%\bw 
The first result states that a decimated long-range Gibbs measure in one dimension becomes non-Gibbsian,
%\ew,
in the same vein as comparable results in higher dimensions for short-range models.\\
  The second result addresses  the behaviour of such models under  inhomogeneous fields, in particular  
  %{\bf correlated,} 
  external fields which decay to zero  polynomially as 
  $\frac{1}{(|i|+1)^{\gamma}}$.
  %\ew.
   We study how the critical decay power of the field, $\gamma$, for which the phase transition persists and the decay power $\alpha$ of the Dyson model compare, extending recent results for short-range models on lattices and on trees.
We also briefly point out some analogies between these results.

\footnotesize

\vspace{7cm}

 {\em  AMS 2000 subject classification}: Primary- 60K35 ; secondary- 82B20

{\em Keywords and phrases}: Long-range Ising models, hidden phase transitions, generalized Gibbs measures, slowly decaying correlated external fields.

\normalsize
\section{Introduction}
\
%\hspace{.5cm}
%\bw 
In this short review we investigate some properties of one-dimensional long-range spin models, also known as Dyson models. In his original work, Dyson \cite{Dys68} considered an Ising spin system with formal Hamiltonian given by 
\[
H(\omega) = - \sum_{i>j} J(|i-j|)\omega_i\omega_j
\]
and $J(n) \geq 0$ for $n\in \mathbb{N}$ e.g. of the form $J(n)=n^{-\alpha}$. \\
There is no phase transition for this model,   
%either if
 %{\bf 
if the series $M_0=\sum_{n=1}^{\infty} J(n)$ is infinite, since then  there is an infinite energy gap between the ground states and all other states, which yields that at all finite temperatures the system is expected to be ordered. Neither is there a transition  if $|\{n:J(n) \neq 0 \}| < \infty$, by \cite{Rush}, since then the system is 
%expected to be 
disordered at all finite temperatures. See also \cite{BLP} and \cite{CO} for accessible proofs of different versions of  this absence of a transition under conditions of sufficiently fast polynomial decay of $J(n)$. 
%\bw 
Thus  in particular,  there is no phase transition for $J(n)$ being of finite range, and neither for $J(n)=n^{-\alpha}$ with $\alpha > 2$. \\
%\ew
A conjecture due to  Kac and Thompson 
%conjecture
 \cite{Kac}, early on, stated that there should be a phase transition for low enough temperatures if and only if $\alpha \in (1,2]$.
Dyson proved a part of the Kac-Thompson conjecture, namely that for long-range models of the form $n^{-\alpha}$ with $\alpha \in (1,2)$ there is a phase transition. 
%The Kac-Thompson conjecture \cite{Kac} states that there should be a phase transition for low enough temperatures if and only if $\alpha \in (1,2]$. 
Note that for $M_0<\infty$ the infinite-volume measure is well defined.

%\ew 

We will consider, analogously to Dyson, one-dimensional ferromagnetic  models with slowly decaying  pair interactions of the form $ J(|i-j|) = \frac{1}{|i-j|^\alpha}$, for appropriate values of the 
decay parameter, $\alpha \in (1,2]$,  which
display a phase transition at low temperature. This  makes Dyson models  particularly interesting, because they thus can exhibit phase coexistence even in one dimension, which is very unusual. 
Varying this decay parameter plays a similar role as varying the dimension in 
short-range models. This can be done in a continuous manner, so one obtains  analogues of well-defined models in continuously varying non-integer dimensions. This is a major reason why these models have attracted a lot of attention in the study of phase transitions and critical behaviour (see e.g. \cite{CFMP} and references therein).

 In this paper, we  first sketch the proof of the fact that,  at low enough temperature, under a decimation 
transformation the low-temperature measures of the Dyson models are mapped to non-Gibbsian measures. Indeed, similar to what happens for short-range models in higher dimensions, in the phase transition region 
($1 < \alpha \leq 2$ and low enough temperature), decimating the Gibbs measures to half the spins leads to non-Gibbsianness of the 
decimated measures. This is obtained by showing the alternating configuration to be  a point of essential
discontinuity for the (finite-volume) conditional probabilities of the 
decimated Gibbs measures.

Just as with external fields or boundary conditions, the  configuration of renormalised spins, acting on the system of  ``hidden spins" which are to be integrated out, can prefer one of the phases, and there are choices where this preference depends  only on spins far away.  The renormalised spins  can act as some kind of (possibly correlated) random field, acting on the other (hidden) spins.

 We have extended our analysis to consider the effects of more general, possibly decaying, external fields on Dyson models and discuss how Dyson models in external fields decaying to zero 
 %\bw 
 as $\frac{1}{(|i|+1)^{\gamma}}$   behave as regards phase coexistence. Again similarly to what happens in short-range models, it appears that the existence of a plurality of Gibbs measures persists when the decay of the 
field is fast enough, whereas for slowly decaying fields we expect that there is only one Gibbs measure which survives, namely the one favoured by the field. What the appropriate decay parameter of the field, $\gamma$, which separates the two behaviours is,  depends on the Dyson decay parameter $\alpha$.   This extends recent results on short-range models on either lattices or trees.
\medskip
%****

%Moreover we discuss the behaviour of Dyson models in an external field which 
%decays to zero far from the origin. We discuss the question how fast the decay
%of the external field  should be for the phase coexistence to persist. The critical value of this decay power $\gamma$ is compared with the parameter $\alpha$ of the Dyson interaction. 

The review is organized as follows. 
In Section 2, we introduce notations and definitions of  Gibbs measures 
%in mathematical statistical mechanics --  including "global specifications"  \cite{FP} -- and our 
and describe what is known about phase transitions in  Dyson models. 
In Section 3, we introduce the decimation transformation -- a renormalization 
transformation that keeps odd or even spins only -- and sketch how to prove non-Gibbsianness 
at low temperature for the decimated Gibbs measures of the Dyson models. 
%whose interactions are so slowly decaying that, 
We show that, conditioned on the even spins to be alternating,  a  
``hidden phase transition'' occurs in the system of odd spins. 
%Eventually, in Section 4, we extend previous results to show that  this decimated measure is included in the class of Almost Gibbsian measures, and 
%We also comment on some related issues. 
In Section 4 we will discuss Dyson models in decaying fields.
%This paper is based on joint work with Arnaud Le Ny, Rodrigo Bissacot,Eric Endo, Bruno Kimura and Wioletta Ruszel\cite{ELN} and on {BEEKR}.
\section{ Gibbs Measures and Dyson Models}
\subsection{Specifications and Measures}

We refer to \cite{ELN} and \cite{BEEKR} for proofs and  more details on the general formalism considered here. \\
Dyson models  are  ferromagnetic Ising models 
with   long-range pair-interactions in one dimension, possibly with an external field which we will take possibly  inhomogeneous, random  and{/}or correlated. \\
%acting as self-interactions.
 We study these models  within a  more general class of lattice (spin) models with Gibbs measures on infinite-volume product configuration spaces $(\Omega,\mathcal{F},\rho)=(E^\Z,\mathcal{E}^{\otimes \Z},\mu_o^{\otimes Z})$, the single-site state space being the Ising space 
%$(E,\mathcal{E},\rho_0)$ with 
$E=\{-1,+1\}$,
%$\mathcal{E}=\mathcal{P}(\{-1,+1\})$ and
with the a priori counting 
measure $\mu_0=\frac{1}{2} \delta_{-1} + \frac{1}{2} \delta_{+1}$. We denote by 
$\mathcal{S}$ the set of the finite subsets of $\Z$ and, for any $\La \in \s$, 
write $(\Om_\La,\mathcal{F}_\La,\rho_\La)$ for the finite-volume configuration 
space $(E^\La,\mathcal{E}^{\otimes \La},\mu_o^{\otimes \La})$ -- and extend afterwards the notations when considering infinite subsets $S \subset \Z$ and (restricted) infinite-volume configuration spaces
$(\Omega_S,\mathcal{F}_S, \mu_S) \ni \sigma_S$.

Microscopic states or configurations, denoted by $\si,\om, \eta,\tau,\;$, etc., are  elements of  $\Omega$ 
equipped with the product topology of the discrete topology on $E$ for which these configurations are close when they coincide on large finite regions $\Lambda$ (the larger the region, the closer). For $\omega \in \Om$, a  neighborhood base is  provided by 
% $\big(\mathcal{N}_\Lambda(\omega)\big)_{\Lambda \in \mathcal{S}}$ with, for any $\Lambda \in \s$,
$$
\mathcal{N}_L(\omega)=\Big \{ \sigma \in \Omega : \sigma_{\Lambda_L}=\omega_{\Lambda_L},\; \sigma_{\Lambda_L^c} \; {\rm arbitrary} \Big\},\;L \in \N,\;  \Lambda_L:=[-L,+L] \in \mathcal{S}.
$$
%{\bf Arnaud : in the way we wrote things, we might not need to introduce this notation.

For any integers $N > L$, we shall also consider particular open subsets of neighborhoods 
%$\mathcal{N}_{N,L}(\omega)$ on which the configuration is $+$ (resp. $-$) on an annulus $\Delta \setminus \Lambda$ for $\Delta \supset \Lambda$, defined for all $\Lambda \in \s,\; \om \in \Om$ as
$$
\mathcal{N}_{N,L}^+(\omega) = \Big \{ \sigma \in \mathcal{N}_L(\omega) : \sigma_{\Lambda_N \setminus \Lambda_L}  = +_{\Lambda_N \setminus \Lambda_L},\; \sigma\; {\rm arbitrary \; otherwise} \Big\} \; \big({\rm resp.} \; \mathcal{N}_{N,L}^-(\omega)\big).
$$
% = \Big \{ \sigma \in\mathcal{N}_\Lambda(\omega) : \sigma_{\Delta \setminus \Lambda } = -_{\Delta \setminus \Lambda},\; \sigma_{\Delta^c} \; {\rm arbitrary} \Big\}.
%\end{eqnarray*}
%  For this 
%topology, continuous functions coincide with  quasilocal functions.
%, that is, uniform limits of local functions, 
%the latter being $\mathcal{F}_\La$-measurable functions for some $\La \in \s$. A
 %function is said to be {\em right-continuous} (resp. {\em left-continuous}) 
%when for every $\om \in \Om$, $\lim_{\La \uparrow \s} f(\om_\La +_{\La^c})=f(\om)$ 
%(resp. $\lim_{\La \uparrow \s} f(\om_\La -_{\La^c})=f(\om))$, where one writes 
%$\om_\La$ for its projection on $\Om_\La$, and $+$ (resp. $-$) for the configurations
 %whose value are respectively  $+1$ (resp. $- 1$) everywhere. We also generically 
%When considering infinite subsets $S \subset \Z$, 
%for which 
%all the preceding notations 
%defined for finite $\Lambda$ extend naturally 
%($\Omega_S,\mathcal{F}_S, \mu_S, \sigma_S$, etc.). 
%Important events to be 
%considered are the {\em asymptotic events}, which are the elements of the 
 %tail $\sigma$-algebra $\mathcal{F}_{\infty}=\cap_{\Lambda \in \mathcal{S}} \mathcal{F}_{\Lambda^c}$. These events typically do not depend on any local  behaviour, that is, they are insensitive to changes of any finite number of spins, and are mostly obtained by some limiting procedure. 
We denote by $C(\Om)$ the set of continuous  (quasilocal) functions on $\Om$,  characterized by
%In our  finite state-space set-up, continuity is equivalent to uniform continuity and to   quasilocality\footnote{Continuous functions are uniform  limits of local functions, explaining the terminology {\em quasilocal} \cite{Fer,HOG}.}, so that one has
\be \label{qlocfu} 
f \in C(\Omega) \; \Longleftrightarrow \; \lim_{\Lambda \uparrow \s} \sup_{\sigma,\omega:\sigma_\Lambda=\omega_\Lambda} \mid f(\omega) - f(\sigma) \mid = 0.
\ee
Monotonicity for functions and measures concerns the natural partial (FKG) order "$\leq$ ", which we have on our Ising spin systems : $\sigma \leq \omega$ if and only if 
$\sigma_i \leq \omega_i$ for  all  $i \in \Z$. Its maximal and minimal elements 
are the configurations $+$ and $-$, and this order extends to functions: 
$f:\Omega \longrightarrow \mathbb{R}$ is called {\em monotone increasing}  when 
$\sigma \leq \omega$ implies $f(\sigma) \leq f(\omega)$. For measures,  we write $\mu \leq \nu$ if and only
if $\mu[f] \leq \nu[f]$ for all $f$ monotone increasing\footnote{We denote 
$\mu[f]$ for the expectation $\E_\mu[f]$ under a measure $\mu$.}.
\\

Macroscopic states are  represented by probability measures  
on  $(\Omega,\mathcal{F},\rho)$, whose main description -- at least in mathematical statistical mechanics --  is  in terms of consistent systems of regular versions of finite-volume conditional probabilities with prescribed boundary conditions, within the so-called {\em DLR formalism} \cite{Dob1,LaR}.  To do so, one introduces families of probability kernels that are natural candidates to represent such versions of conditional probabilities. 
\begin{definition}[Specification]:\\
A {\em specification} $\gamma=\big(\gamma_\Lambda\big)_{\Lambda \in \s}$  on $(\Omega,\mathcal{F})$ is a family of probability kernels  $\gamma_\Lambda : \Omega_\Lambda \times \mathcal{F}_{\Lambda^c} \; \longrightarrow \; [0,1];\; (\omega,A) \; \longmapsto \;\gamma_\Lambda(A \mid \omega)$
s.t. for all $\Lambda \in \mathcal{S}$:
\begin{enumerate}
%\item For all $\omega \in \Omega$, $\gamma_\Lambda(\cdot | \omega)$ is a probability measure on $(\Omega,\mathcal{F})$.
%\item For all $A \in \mathcal{F}$, $\gamma_\Lambda(A | \cdot)$ is $\mathcal{F}_{\Lambda^c}$-measurable.
\item (Properness) For all $\omega \in \Omega$, $\gamma_\Lambda(B|\omega)=\mathbf{1}_B(\omega)$ when $B \in \mathcal{F}_{\Lambda^c}$.
\item (Finite-volume consistency) For all $\Lambda \subset \Lambda' \in \s$, $\gamma_{\Lambda'} \gamma_{\Lambda}=\gamma_{\Lambda'}$ where 
\be \label{DLR0}
\forall A \in \mathcal{F},\; \forall \omega \in \Omega,\;(\gamma_{\Lambda'} \gamma_\Lambda)(A | \omega)=\int_\Omega \gamma_\Lambda(A | \sigma) \gamma_{\Lambda'}(d \sigma | \omega).
\ee
\end{enumerate}
\end{definition}
These kernels also act on functions and on measures: for all $f \in C(\Omega)$ or $\mu \in \mathcal{M}_1^+$,
$$
\gamma_\Lambda f(\omega):=\int_\Omega f(\sigma) \gamma_\Lambda (d \sigma | \omega)=\gamma_\Lambda [f | \omega] \; {\rm and} \; 
\mu \gamma_\Lambda [f] : = \int_\Omega (\gamma_\Lambda f)(\omega) d \mu (\omega)= \int_\Omega \gamma_\Lambda [f | \omega] \mu(d \omega).
$$

%These objects are designed to represent consistent systems of conditional probabilities, with the important 
%objection 
%additional property
%that they are defined everywhere and not only almost-surely as ordinarily conditional probabilities would have been required to be. However, as we do not have a measure to begin with, the notion of "almost surely" a priori does not make sense.   

%For a given specification, different measures can then have their conditional probabilities represented by the  same specification (and satisfy the {\em DLR equations} (\ref{DLR1})) but live on different full-measure sets. This leaves the door open to a mathematical description of phase transitions, which is well known, in particular  for the ferromagnetic (n.n.) Ising model on the square lattice $\Z^2$ \cite{Grif}, but also for our long-range Ising models on $\Z$.

\begin{definition}[DLR measures]:\\
A probability measure $\mu$ on $(\Omega,\mathcal{F})$ is said to be consistent with a specification $\gamma$ (or specified by $\gamma$) when  for all $A \in \mathcal{F}$ and $\Lambda \in \s$
\be \label{DLR1}
\mu[A|\mathcal{F}_{\Lambda^c}](\omega)=\gamma_\Lambda(A|\omega), \; \mu{\rm -a.e.} \;  \omega.
\ee
%Equivalently, $\mu$ is consistent with $\gamma$ when $\mu=\mu \gamma_\Lambda$ for all $\Lambda \in \s$, i.e. when
%\be \label{DLR2}
%\int (\gamma_\Lambda f) d \mu = \int f d \mu,\forall \Lambda \in \s,\; \forall f \in \mathcal{F}_{\rm{loc}}
%\ee
%or, in an even shorter form, if and only if 
%\be \label{DLR3}
%\forall \Lambda \in \s,\; \mu \gamma_\Lambda = \mu
%\ee
We denote by $\mathcal{G}(\gamma)$ the set of measures consistent with $\gamma$. 
%For a translation-invariant specification, $\mathcal{G}_{\rm{inv}}(\gamma)$ is the set of translation-invariant elements of $\mathcal{G}(\gamma)$.
\end{definition}

 A {\it specification}  $\gamma$ is said to be  {\bf quasilocal} when  for any local function $f$, the image $\gamma_\Lambda f$ should be a continuous function of the boundary condition :
\be\label{qlocmes}
\gamma \; {\rm quasilocal} \; \; \Longleftrightarrow \; \; \gamma_\Lambda f \in C(\Omega) \; {\rm for \; any} \; f \; {\rm local} \;  ({\rm or \; any} \; f\;{\rm  in}\; C(\Omega)).
\ee
A {\it measure} is said to be  quasilocal when it is specified by a quasilocal specification. 

A particularly important subclass of quasilocal measures consists of the {\em Gibbs measures} with (formal) Hamiltonian $H$ defined  via a potential $\Phi$, a family $\Phi=(\Phi_A)_{A \in \s}$ of local functions $\Phi_A \in \mathcal{F}_A$.  The contributions of spins in finite sets $A$ to the total energy define the {\em finite-volume  Hamiltonians with free boundary conditions} 
\be \label{Ham}
\forall \Lambda \in \s,\; H_\Lambda(\omega)=\sum_{A \subset \Lambda} \Phi_A(\omega),\; \forall \omega \in \Omega.
\ee
To define Gibbs measures, we require for $\Phi$ that it is  {\em Uniformly Absolutely Convergent} (UAC), i.e. 
that $\sum_{A \ni i} \sup_\omega |\Phi_A(\omega)| < \infty, \forall i \in \Z$.
%\be \label{UACPot}
%\forall i \in \Z,\;\sum_{A \ni i} \sup_\omega |\Phi_A(\omega)| < \infty.
%\ee
%For such a potential, 
One then can give sense to the  {\em Hamiltonian at volume $\Lambda \in \s$ with boundary condition $\omega$} defined for all $\sigma,\omega \in \Om$ as $H_\Lambda^\Phi(\sigma | \omega) := \sum_{A \cap \Lambda \neq \emptyset} \Phi_A(\sigma_\Lambda \omega_{\Lambda^c}) (< \infty)$.
%\be \label{Hambc}
%H_\Lambda^\Phi(\sigma | \omega) := \sum_{A \cap \Lambda \neq \emptyset} \Phi_A(\sigma_\Lambda \omega_{\Lambda^c}) (< \infty).
%\ee
The {\em Gibbs specification at inverse temperature $\beta>0$} is then defined by
\be \label{Gibbspe}
\gamma_\Lambda^{\beta \Phi}(\sigma \mid \omega)=\frac{1}{Z^{\beta \Phi}_\Lambda(\omega)} \; e^{-\beta H_\Lambda^\Phi(\sigma | \omega)} (\rho_\Lambda\otimes \delta_{\omega_{\Lambda^c}}) (d \sigma)
\ee
where the partition function $Z_\Lambda^{\beta \Phi}(\omega)$ is an important normalizing constant. Due to the, in fact rather  strong, UAC condition, these specifications are quasilocal. It appears that the converse is also true up to a non-nullness condition\footnote{expressing that $\forall \Lambda \in \s,\; \forall A \in \mathcal{F}_\Lambda$, $\rho(A)>0$ implies that $\gamma_\Lambda (A | \omega) >0$ for any  $\omega \in \Om$.} (see e.g. \cite{HOG, Fer, Ko, Su, ALN2}) and one can take :
\begin{definition}[Gibbs measures]:\\
$\mu \in \mathcal{M}_1^+$ is a Gibbs measure iff $\mu \in \mathcal{G}(\gamma)$,  where $\gamma$ is a non-null and quasilocal specification.
\end{definition}

Quasilocality, called {\em Almost Markovianness} in \cite{Su}, is a natural way to extend the global (two-sided) Markov property. When $\mu \in \mathcal{G}(\gamma)$ is quasilocal, then for any  $f$ local and  $\Lambda \in \s$,  the conditional  expectations of $f$ w.r.t. the outside of $\Lambda$ are $\mu$-a.s. given by $\gamma_\Lambda f$, by  (\ref{DLR0}), and each conditional probability has a version which itself is a continuous function of the boundary condition, so one gets for any $\omega$
\be \label{esscont}
\lim_{\Delta \uparrow \mathbb{Z}} \sup_{\omega^1,\omega^2 \in \Omega}  \Big| \mu \big[f |\mathcal{F}_{\Lambda^c} \big](\omega_\Delta \omega^1_{\Delta^c}) - \mu \big[f |\mathcal{F}_{\Lambda^c} \big](\omega_\Delta\omega^2_{\Delta^c})\Big|=0
\ee
%which yields an (almost-sure) asymptotically  weak dependence on the conditioning.
%,
%which can be seen as an extended  Markov property. 
Thus, for Gibbs measures the conditional probabilities always have continuous versions, or equivalently
%In particular, for Gibbs measures, it is not possible to change  its conditional probabilities on exceptional sets in order to get a discontinuous version: one says that 
 there is no point of essential discontinuity. Those are configurations which are points of discontinuity for ALL versions of the conditional probability. In particular one cannot make conditional probabilities continuous by redefining them on a measure-zero set if such points exist. In the generalized Gibbsian framework, one also says that such a configuration is a {\em bad configuration} for the considered measure, see e.g. \cite{ALN2}.
The existence of such bad configurations implies non-Gibbsianness of the associated measures.

\subsection{Dyson models: Ferromagnets in One Dimension}

%In our  framework\footnote{Or more generally when the configuration space is {\em standard Borel}, see \cite{HOG}.}, for any given $\mu \in \mathcal{M}_1^+$, it is always possible to construct a specification $\gamma$ such that $\mu \in \mathcal{G}(\gamma)$ (see e.g.  Goldstein \cite{Gold}, Preston \cite{Pr} or Sokal \cite{Sok}). Nevertheless, even in such a framework, there exist specifications $\gamma$ for which $\mathcal{G}(\gamma)=\emptyset$ (see e.g. \cite{HOG, ALN2}), others where $\mathcal{G}(\gamma)=\{\mu\}$ but also -- and this is more interesting for us -- some for which this set contains more than one element. In the latter case ,
% we say in mathematical statistical mechanics that 
% there is a {\em phase transition}. The set of DLR measures is then known to be a convex set whose extremal elements are trivial on the tail $\sigma$-algebra $\mathcal{F}_\infty$. Any other element of $\mathcal{G}(\gamma)$ admits a unique\footnote{It is a {\em Choquet simplex}, see \cite{Dy, HOG}.} convex combination of the extremal elements and is characterized by its action on the tail $\sigma$-algebra $\mathcal{F}_\infty$ \cite{VEFS, HOG}. We focus here on such a case in dimension one:

\begin{definition}[Dyson models]:\\
Let $\beta >0$ be the inverse temperature  and consider $1 < \alpha \leq 2$. We call 
%{\em Dyson-Ising specification} 
a {\em Dyson model with decay parameter $\alpha$} the Gibbs specification (\ref{Gibbspe}) with pair-potential $\Phi^D$ defined  for all $\omega \in \Omega$ by

\be \label{Dys}
\Phi_A^D(\omega)= - \frac{1}{|i-j|^\alpha} \om_i \om_j \; {\rm when} \; A=\{i,j\} \; \subset \mathbb{Z},\; {\rm and}\; \Phi_A^D \equiv 0 \; {\rm otherwise}.
\ee

%\be \label{LRDysonSpe}
%\gamma_\La^D(d \si | \om) = \frac{1}{Z_\La^\beta(\om)} \; e^{\beta \sum_{i \neq j, i \in \La, j \in \Z} \frac{1}{|i-j|^\alpha} \si_i \si_j} \; \rho_\La \otimes \delta_{\om_{\La^c}} (d \si)
%\ee
%where the normalization $Z_\La^\beta(\om)$ is the usual partition function.

We shall also  consider Dyson models with non-zero magnetic fields $h=(h_i)_{_
i \in \Z}$ acting as an extra self-interaction part
$
\Phi_A^D(\omega)= - h_i \omega_i  \; {\rm when} \; A=\{i\}  \; \subset \mathbb{Z}
$
\end{definition}

We first use that as a consequence of FKG property \cite{FKG,Hul}, the Dyson specification is {\em monotonicity-preserving} 
\footnote{in the sense that for all bounded increasing functions 
$f$, and  $\La \in \s$, the function $\gamma_\La^D f$ is 
increasing.},  
%to state 
which implies
that  the weak limits obtained by 
using as boundary conditions the maximal and minimal elements of the
order $\leq$ are well defined and are the extremal elements of $\mathcal{G}(\gamma^D)$.\\

% Indeed, one can learn in e.g. \cite{FP,Hul,Leb}  that
\begin{proposition}\label{Wlimit}\cite{FP,Hul,Leb}:
  For $\alpha >1$ ( and not only for  $ \alpha \in (1,2]$), the weak limits
\be \label{muplusminus}
\mu^-(\cdot) := \lim_{\La \uparrow \mathbb{Z}} \gamma_\La^D (\cdot | -)\; \; {\rm and} \; \; \mu^+(\cdot) := \lim_{\La\uparrow \mathbb{Z}}  \gamma_\La^D (\cdot | +)
\ee
are well-defined, translation-invariant and extremal elements of $\mathcal{G}(\gamma^D)$. For any $f$ bounded increasing, any other measure $\mu \in \mathcal{G}(\gamma^D)$ satisfies
\be \label{stochdom}
\mu^-[f] \leq \mu[f] \leq \mu^+[f].
\ee
Moreover, $\mu^-$ and $\mu^+$  are respectively left-continuous and right-continuous.
\end{proposition}

 While $\mu^-$ and $\mu^+$ coincide at high temperatures, and at all temperatures when there is  fast decay,  $\alpha>2$, one main peculiarity of this one-dimensional model is  thus that when the range is long enough ($1<\alpha \leq 2$), it is possible to recover  low-temperature behaviours usually 
%devoted to 
 associated to
higher dimensions for the standard Ising model, in particular 
%what is generically called a phase transition in mathematical statistical mechanics. 
phase transitions can occur.
For more details on the history of the proofs, one can consult \cite{ELN} and references therein or below.

\begin{proposition}\label{DyFrSp} \cite{Dys,frsP,Rue72,HOG,FV,FILS,ACCN,CFMP,LP,Joh}.

The Dyson model with potential (\ref{Dys}), for $1< \alpha \leq 2$, exhibits a {\em phase transition at low temperature}:
$$
\exists \beta_c^D >0, \; {\rm such \; that} \; \beta > \beta_c^D \; \Longrightarrow \; \mu^- \neq \mu^+ \; {\rm and} \; \mathcal{G}(\gamma^D)=[\mu^-,\mu^+]
$$
where the extremal  measures $\mu^+$ and $\mu^-$ are translation-invariant. They have in particular opposite magnetisations   $\mu^+[\sigma_{0}]=-\mu^-[\sigma_{0}]=M_0(\beta, \alpha)>0$ at low temperature. Moreover, the Dyson  model in a non-zero homogeneous field $h$ has a unique Gibbs measure. 

\end{proposition}

%{\bf Proofs, where to find them:} \\
%The existence of phase transitions at low temperature comes was first proved by  Dyson for $1<\alpha<2$ \cite{Dys} and  Fr\"ohlich/Spencer for $\alpha=2$ \cite{frsP}. \\
%Uniqueness in non-zero field follows immediately from a theorem given in the Appendix of \cite{Rue72} which applies to all ferromagnetic Ising pair interactions, including Dyson models. The proof  uses the Lee-Yang  circle theorem to obtain an analyticity property of the pressure, as well as the FKG stochastic domination. See also \cite{HOG}, Notes to Chapter 16.2,  or the detailed proof of \cite{FV} in the standard Ising case.
%Later proofs of phase transitions used Reflection Positivity \cite{FILS}, the Fortuin-Kasteleyn  random cluster representation \cite{ACCN}, or contour-type arguments\cite{CFMP,LP}, see also \cite{Joh}. 

We remark that the infinite-volume limit of a state (or a magnetisation) in which there is a $+$ (resp. $-$)-measure or a Dyson model in a field $h >0$ (resp. $h<0$) outside some interval
%the boundaries
  is the same 
  % $+M_0(\alpha,\beta)$  (resp. $-M_0(\alpha, \beta)$) 
  as that obtained from $+$ (resp. $-$)-boundary conditions (independent of the magnitude of $h$).  This can be e.g. seen by an extension of the arguments of \cite{LebP}, see also \cite{Leb2}. Notice that taking the $+$-measure of the zero-field Dyson model outside a finite volume enforces this same measure inside (even before taking the limit); adding a field makes it it more positive, and taking the thermodynamic limit then recovers the same measure again.

The case of $\alpha=2$ is more complicated to analyse, and richer in its behaviour, than the other ones. 
There exists a hybrid transition (the "Thouless effect"), as the magnetisation is discontinuous while the energy density is continuous at the transition point. 
Moreover, there is second transition below this transition temperature. In the intermediate phase there is a positive magnetisation with non-summable covariance, while at very low temperatures the covariance decays at the same rate as the interaction, which is summable. For these results, see  \cite{ACCN, I,IN},  and  also the more recent description in \cite{LP}.

\section{Decimation}

We first apply a decimation transformation to the lattice $2 \Z$.
Similarly to what was discussed in \cite{VEFS}, to analyse whether the transformed measure is a Gibbs measure, and in particular to show that it is non-Gibbsian,  we have to show that conditioned on  a particular configuration of the transformed spins, the "hidden spins" display a phase transition. If we choose  this particular configuration to be the alternating one, each hidden spin feels opposite terms from the left and the right side,  coming from all odd distances. Thus the conditioned model is a Dyson model in zero field, at a reduced temperature. As such it has phase transition. \\
To translate this hidden phase transition into nonlocality of the "visible" transformed spins follows straightforwardly the arguments of \cite{VEFS}.
See \cite{ELN} for the details. We make use of the fact that one can define global specifications, so there are no measurability problems due to global conditioning.

We start from $\mu^+$ (\ref{muplusminus}), the $+$-phase of a Dyson model  without external field in the phase transition region 
%(with slow decay $1 < \alpha \leq 2$ at low temperature) 
and apply the {\em decimation transformation}
\be \label{DefDec}
 T \colon (\Omega,\mathcal{F})  \longrightarrow (\Omega',\mathcal{F}')=(\Omega,\mathcal{F}); \; 
\omega \; \;   \longmapsto \omega'=(\omega'_i)_{i \in
\mathbb{Z}}, \; {\rm with} \;  \omega'_{i}=\omega_{2i}
\ee

Denote $\nu^+:=T \mu^+$ the decimated $+$-phase, formally defined as an image measure via
$$
\forall A' \in \mathcal{F'},\; \nu^+(A')=\mu^+(T^{-1} A')=\mu^+(A) \; {\rm where} \; A=T^{-1} A'= \big\{\omega: \omega'=T (\omega) \in A' \big\}.
$$
%When necessary, we distinguish between original and image sets using  primed notation

%\footnote{Notice that by  rescaling  the configuration spaces $\Omega$ (original) and $\Omega'$ (image) are identical.}. 

%In particular, we get the following expressions in terms of global specifications first, expressed themselves in terms of  constrained measures afterwards, of conditional expectations of the spin at the origin, used next section to prove an essential discontinuity. \\

%****************************************************************\\
%No changes for the moment for the following paragraph. Maye we should check at the end that things are not told too many times in different forms...\\
%***********************************************************\\

We  study the continuity of  conditional expectations
%, under $\nu^+$, 
under decimated Dyson Gibbs measures
of the spin at the origin when the outside is fixed 
%under 
 in
some special configuration  $\omega'_{\rm alt}$. By definition,
%As the conditioning takes place on an infinite set with infinite complement, we need here  global specifications for the decimated measures built as in Theorem \ref{globspe} got from  \cite{FP}.
% or \cite{ALN}. To build these specifications, we first note that
%On the other hand, if in the "annulus" we impose all plus spins for the primed sites, our magnetisation at the origin will be larger than that of the plus state of the constrained system.  In fact, due to the fact that the Dyson model in a positive external field has only one Gibbs measure, the influence transmitted through the annulus decays with the distance to the external boundary, whatever the boundary condition. A similar argument applies to the situation in which the primed spins in the annulus are all minus.  
%{\bf REMARK, the nonGibbsianness can be proven at temepratures strictly below the phase transition temperature of %the Dyson models (the constrained model has a lower transition temperature)}
%To do so, consider a basis of neighborhood $(\mathcal{N}_{\Lambda'}(\omega'_{\rm alt})_{\Lambda' \in \mathcal{S}}$  %for $\Lambda' \in \s$ containing the origin, and any $\omega' \in \mathcal{N}_{\Lambda'}(\omega'_{\rm alt})$, we %express the conditional expectations  using the very definition of $\nu^+$ as an image measure of $\mu^+$ via the %decimation transformation $T$: for any  $\omega' \in \mathcal{N}_{\Lambda'}(\omega'_{\rm alt})$ and any  $\omega \in %T^{-1} \{\omega'\}$,
\be \label{condmagn}
 \nu^+[\sigma'_0 | \mathcal{F}_{\{0\}^c} ](\omega') = \mu^+[\sigma_0 | \mathcal{F}_{S^c} ](\omega),\; \nu^+{{\rm -a.s.}}
\ee
where $S^c=(2 \mathbb{Z}) \cap \{0\}^c$, i.e. with $S= (2 \mathbb{Z})^c \cup \{0\}$ is not  finite: {\em the conditioning  is  not on the complement of a finite set},  and although the extension of the DLR equation to infinite sets is direct in case of uniqueness of the DLR-measure for a given  specification \cite{FP, Foll, Gold2},  it can be more problematic otherwise: it is valid for finite sets only and  measurability problems might arise in case of phase transitions  when one wants to extend them to infinite sets. Nevertheless, beyond the uniqueness case, such an extension was made possible by Fern\'andez and Pfister  \cite {FP} in the case of attractive models.
 As we will make essential use of it, we describe it now in our particular case. 
The concept they introduced is that of  a {\em global specification}, and this is in fact a central tool in some of our arguments.

\begin{definition}[Global specification \cite{FP}]\label{Glob}:\\
A {\em global specification} $\Gamma$ on $\Z$ is a family of probability kernels $\Gamma=(\Gamma_S)_{S \subset \Z}$ on $(\Omega_S,\mathcal{F}_{S^c})$ such that for {\em any} $S$ subset of $\Z$:
\begin{enumerate}
\item $\Gamma_S(\cdot | \omega)$ is a probability measure on $(\Omega,\mathcal{F})$ for all $\omega \in \Omega$.
\item $\Gamma_S(A | \cdot)$ is $\mathcal{F}_{S^c}$-measurable for all $A \in \mathcal{F}$.
\item $\Gamma_S(B|\omega)=\mathbf{1}_B(\omega)$ when $B \in \mathcal{F}_{S^c}$.
\item For all $S_1 \subset S_2 \subset \Z$, $\Gamma_{S_2} \Gamma_{S_1}=\Gamma_{S_2}$ where the product of kernels is made as in (\ref{DLR0}).
\end{enumerate}
\end{definition}
Similarly to the consistency with a (local) specification, one introduces the {\em compatibility of measures with a global specification}.
\begin{definition}
Let $\Gamma$ be a global specification. 
We write $\mu \in \mathcal{G}(\Gamma)$, or say that  $\mu \in \mathcal{M}_1^+$ is $\Gamma${\em -compatible}, if for all $A \in \mathcal{F}$ and {\em any} $S \subset \Z$,
\be \label{DLR4}
\mu[A|\mathcal{F}_{S^c}](\omega)=\Gamma_S(A|\omega), \; \mu{\rm -a.e.} \;  \omega.
\ee
\end{definition}

Note, by considering $S=\mathbb{Z}$, that $\mathcal{G}(\Gamma)$ contains at most one element. \\

In the case considered here, we get a global specification $\Gamma^+$ such that $\mu^+ \in \mathcal{G}(\Gamma^+)$,  with $S= (2 \mathbb{Z})^c \cup \{0\}$ consisting of the {\em odd integers plus the origin}.  Hence $S =(2 \mathbb{Z})^c \cup \{0\}$ and (\ref{condmagn}) yields for 
$\nu^+$-a.e.
 all
 $\omega' \in \mathcal{N}_{\Lambda'}(\omega'_{\rm alt}) $ and $\omega \in T^{-1} \{\omega'\}$: 
\be \label{condmagn2}
 \nu^+[\sigma'_0 | \mathcal{F}_{\{0\}^c} ](\omega') = \Gamma_{S}^+ [\sigma_0 | \omega] \; \; \mu^+{\rm -a.e.} (\omega). 
\ee
to eventually get (see \cite{FP,ELN}) an expression of the latter in terms of a constrained measure $\mu^{+,\omega}_{(2\mathbb{Z})^c \cup \{0\}}$, with $\omega \in T^{-1} \{\omega'\}$  so that we get for any 
%{\bf ( ? $\nu^+$-a.e ?)} 
$\omega' \in \mathcal{N}_{\Lambda'}(\omega'_{\rm alt})$,
$$
\nu^+[\sigma'_0 | \mathcal{F}_{\{0\}^c} ](\omega') = \mu^{+,\omega}_{(2\mathbb{Z})^c \cup \{0\}} \otimes \delta_{\omega_{2\mathbb{Z} \cap \{0\}^c}} [\sigma_0].
$$
Thanks to monotonicity-preservation, the constrained measure is explicitly built as the weak limit  obtained by $+$-boundary conditions fixed after a freezing the constraint to be 
%{\bf in} 
$\omega$ on the even sites :
\be \label{constrLimit}
\forall \omega' \in \mathcal{N}_{\Lambda'}(\omega'_{\rm alt}), \forall \omega \in T^{-1} \{\omega'\},\;   \mu^{+,\omega}_{(2\mathbb{Z})^c \cup \{0\}} (\cdot) =\lim_{I \in\s,I \uparrow (2 \mathbb{Z})^c  \cup \{0\}} \gamma^D_I (\cdot\mid +_{(2 \mathbb{Z})^c  \cup \{0\})} \omega_{2 \mathbb{Z} \cap\{0\}^c}).
\ee

Observe that when a  phase transition holds for the Dyson specification -- at low enough T for $1 < \alpha \leq 2$ -- the same is true for the constrained specification 
%(\ref{const}) 
{\em with alternating constraint} (although at a lower T). This   phase transition then implies  non-Gibbsianness of $\nu^+$ (and for all other Gibbs measures of the model, see \cite{ELN}).   
% We shall come back to this later, before we state and prove our main result. 

%\subsection{Non-Gibbsianness at Low Temperature}
\begin{theorem}\label{thm2} (\cite{ELN}) :
 Let $\alpha \in (1,2]$, let $\mu$ be a Gibbs measure for the interaction given by \eqref{Dys} and let the transformation $T$ be defined by \eqref{DefDec}. Then for  low temperatures, $\beta > 2^{\alpha}\beta^D_{c}$, the decimated measure $\nu = T \circ \mu$ is non-Gibbs. 
%For any  $1<\alpha \leq 2$, at low enough temperature
% $\beta > \beta_c^D$, 
%the decimation  $\nu$ of any 
%$+$-phase
 %Gibbs measure $\mu$ of the Dyson-Ising model, $\nu=T \mu$ is non-Gibbs.
\end{theorem}

\noindent
{\bf Sketch of Proof:}
 The main idea is to prove that the alternating configuration is an essential point of discontinuity for the decimated conditional expectations.  
 As already observed, because any non-fixed site at all odd distances has a positive and a negative spin whose influences cancel, conditioning by this alternating configuration yields a constrained model that is again  a Dyson model at zero field, but at a temperature which is higher by $2^{\alpha}$. This  again has a low-temperature transition in our range of decays $1 < \alpha \leq 2$.
 The coupling constants are  indeed multiplied by a factor $ 2^{- \alpha}$, due to only even distances occurring between interacting (hidden) spins.
 
 % Thus the argument will only work if the temperature is at least  smaller by that factor than the transition temperature of the original Dyson model. 
 
  To prove non-Gibbsianness in \cite{ELN}, we essentially follow the proof strategy  sketched in \cite{VEFS}, by showing that  within a neighborhood $\mathcal{N}_L(\omega'_{{\rm alt}})$, there exists two subneighborhoods $\mathcal{N}_{N,L}^\pm$ of positive measure on which the conditional magnetizations defined on $\mathcal{N}_N,L^\pm$
 \begin{equation}\label{CondMagn}
 M^+ = M^+(\omega)=
\mu^{+,\omega^+}_{(2\mathbb{Z})^c \cup \{0\}}[\sigma_0] \; {\rm and} \; M^- = M^-(\omega)=\mu^{+,\omega^-}_{(2\mathbb{Z})^c \cup \{0\}}[\sigma_0].
\end{equation}
 
 differ significantly.
 
The role of the ``annulus'' where configurations are constrained to be either $+$ or $-$ is played by two large intervals $[-N,-L-1]$ and $[L+1,N]$. Due to the long range of the interaction, their might be a direct influence from the boundary beyond the annulus, to the central interval. To avoid effects from this influence, we take $N$ {\em much} larger than $L$. An argument based on  "equivalence of boundary conditions" as in e.g. \cite{BLP} , under a choice $N =L^{\frac{1}{\alpha-1}}$ then implies that  (\ref{CondMagn}) does hardly  depend on $\omega$.

% holds and modifications beyond annulii do not affect the choice made within it.

Once this choice of big annulus is made, observe that if  we constrain the spins in these two intervals to be either  $+$ or $-$, within these two intervals the measures on the unfixed spins are close to those of the Dyson-type model in a positive, c.q. negative, magnetic field. As those measures are unique Gibbs measures, no influence from the boundary can be transmitted. 
%by 
 %via this ``annulus'', as there is no long-range order.

% Due to the long range of the Dyson interaction, there may be also a direct influence from the boundary, that is from  beyond the annulus, to the central interval, however. But by choosing $N(L)$ large enough -- e.g.  $N =L^{\frac{1}{\alpha-1}}$  -- we can make this direct  influence as small as we want. The special configuration chosen, as indicated above, is   an  alternating one (just as in \cite{VEFS}). 

%Conditioned on all primed spins being alternating \bw$\omega'_{alt}$ \ew, the conditioned model is a Dyson-like model in zero field, due to cancellations, so that a phase transition occurs at low temperature, making it possible to select the phase by boundary conditions arbitrarily far away.

 Indeed, in contrast to the case of the purely alternating configuration, in the case when we condition on all primed spins to be $+$ (resp. $-$) in these large annuli, there is no phase transition and  the system of unprimed spins 
has a unique Gibbs measure.  
It is a Dyson model, again at a heightened temperature, but now in a homogeneous external field,   with positive (resp. negative) magnetisation $+M_0(\beta,\alpha)>0$ (resp. $-M_0(\beta,\alpha)<0$), stochastically larger  (resp. smaller) than the zero-field $+$ ( resp. $-$)-measure. 

 In the $-$-case, in the annulus the magnetisation of the -even-distance- Dyson-Ising model is essentially that of the model with a negative  homogeneous external field $-h$ everywhere, which at low enough temperature and for $L$ large enough is close to (and in fact smaller than) the magnetisation of the Dyson-Ising model under the zero-field $-$-measure, i.e to $-M_0(\beta,\alpha) <0$. 
% (and this $-$-measure is also unique, see \cite{Ker}).  
 Thus the inner interval where the constraint is alternating feels a $-$-like condition from outside its boundary. On the other hand, the magnetisation with the constraint $\omega^+$ will be close to or bigger 
%({\bf keep ?})
than $+M_0(\beta,\alpha)>0$ so that a  non-zero difference is created at low enough temperature. One needs  again to adjust the sizes of $L$ and $N$ to be sure that  boundary effects  from outside the annulus  are negligible  in the inner interval.\\

Thus, for a given $\delta > 0$, e.g. $\delta = \frac{1}{2} M_0(\beta, \alpha)$, 
for  arbitrary $L$ one can find $N(L)$ large enough, such that the expectation of the spin at the origin differs by more than $\delta$. One can therefore  feel the influence from the decimated spins in the far-away annulus,  however large the central interval of decimated alternating spins is chosen. 
Thus,  indeed it holds that
$M^{+} -M^{-} > \delta$,
uniformly in $L$.     \hfill $\square$

In our choice of decimated lattice we 
made use of the fact
%had the advantage
that the constrained system, due to cancellations, again  formed a  
zero-field Dyson-like model. 
%In the case of decimations from $\mathbb{Z}$ to a more diluted lattice $b \mathbb{Z}$ 
%we can do the same, by putting as a constraint alternating intervals of pluses and minuses (or indeed any  configuration of spins which in each even interval alternating with its reflected and spinflipped image. Thus for $b=4$, e.g. the constrained model given as follows $.+--.++-.+--.++-.$ would work).\\
%Other periodic constraints 
%the constrained models 
%could form ferromagnetic models in a periodically varying external field, with zero mean. 
This does not work for decimations to more dilute lattices, but
although  the original proofs of Dyson  \cite{Dys68} and of Fr\"ohlich and Spencer  \cite{frsP}, or the Reflection Positivity proof of \cite{FILS} do no longer apply to such periodic-field cases, the contour-like arguments of \cite{CFMP} and \cite{Joh} could presumably still be modified to include such cases. Compare also \cite{Ker}.

The analysis of \cite{COP} which proves existence of a phase transition for Dyson models in random magnetic fields
for a certain interval of $\alpha$-values should imply that in that case there are many more, random, configurations which all are points of discontinuity. We note that choosing independent spins as a constraint provides a random field which is correlated. However, these correlations decay enough that this need actually not spoil the argument. Similarly, one should be able to  prove that decimation of Dyson models in a weak external field will result in a non-Gibbsian measure.
An interesting question would be to perform the analysis of \cite{MRVS}) or \cite{ALN} to get  a.s. configuration-dependent  correlation decays.

 On the other side of the Gibbs-non-Gibbs analysis, when the range of the interaction is lower, i.e. for $\alpha >2$, or the  temperature is too high, uniqueness holds, for all possible constraints and the transformed measures should be Gibbsian.  Some standard high-temperature results apply, which were already discussed in \cite{VEFS}.  

About  these shorter-range models,  (i.e. long-range models with faster polynomial decay), Redig and Wang \cite{RW} have proved that Gibbsianness was conserved, providing  in some cases ($\alpha >3$) a decay of correlation for the transformed potential. In our longer-range models,  
for intermediate temperatures 
(below the transition temperature but above the transition temperature of the alternating-configuration-constrained model) decimating, both  $+$- and $-$-measures, should imply Gibbsianness, essentially due to the arguments as proposed for short-range models in \cite{HK}.\\
%%%%%%%%%%%%%%%%%%%%%%%%%%%%%%%%%%%%%%%%%%%%%%%%%%%
%%%%%%%%%%%%%%%%%%%%%%%%%%%%%%%%%%%%%%%%%%%%%%%%%%%%%%
\section{Dyson models in decaying fields}

 In this section we consider 
 %in the following section
  one-dimensional Dyson models in a decaying field with decay parameter $\gamma$. 
  %\ew
%If we consider the Dyson model in a polynomially decaying field, with decay parameter $\gamma$, so that the Hamiltonian is
 The corresponding interaction $\Phi^D_{A}(\omega)$ is defined by
\begin{equation}\label{potF}
\Phi^D_A(\omega) = 
\begin{cases}
-\frac{J}{|i-j|^{\alpha}} \omega_i \omega_j& \text{ if } A=\{i,j\} \\
- \frac{h}{ (|i|+1)^{\gamma}} \omega_i & \text{ if } A=\{i\}
\end{cases}
\end{equation}
for some $J,h > 0$. The question raised in \cite{BEEKR} is whether it is possible  to extend results from e.g. \cite{CFMP,COP}, and in particular to  investigate whether and under which conditions the existence of two distinct phases prevails in the presence of an external field. 
Let us mention that one-dimensional Dyson models in a field were considered before, for example in \cite{Ker}, where uniqueness was proven for fields which are either strong enough ($\Phi^D_{\{i\}} = h_i \omega_i $, where there exists $h_0>0$ such that $|h_i| > h_0$) or periodic in large enough blocks. 
 
The main tool we will use are the one-dimensional contours of \cite{CFMP}. Recall that in \cite{CFMP} the authors prove that for $\alpha \in (\alpha^*, 2]$ where $\alpha^*:= 3-\frac{\log(3)}{\log(2)}$ and $h=0$ there exists $\beta^D_{c,0}>0$ such that for all $\beta > \beta^D_{c,0}$
\[
M_0(\beta,\alpha) = \mu^+[\sigma_0] = -\mu^-[\sigma_0] > 0
\]
i.e. there is spontaneous magnetization yielding non-uniqueness of the Gibbs measures, \\ $\mu^+\neq \mu^-$. This result was generalized to all values of $\alpha \in (1,2]$ in \cite{LP}.

Phase coexistence in a positive external field is an unusual phenomenon, since typically Gibbs measures for models in a field are unique. It was previously observed in nearest neighbour pair potentials with polynomially decaying fields in $d\geq 2$, see \cite{BC, BCCP,CV} or for sufficiently fast decaying  (but not necessarily summable) fields on trees \cite{BEE}. In \cite{BCCP} it is proven that in nearest-neighbour models  for $\gamma>1$ and low enough temperatures, there are multiple Gibbs states, whereas for $\gamma < 1$ there is a unique one.  

Pirogov-Sinai is a robust and often applicable version of the Peierls contour argument, applicable in $d \geq 2$, which is the most generally applicable approach in higher dimensions.\\
In \cite{CFMP}, inspired by  and extending results of the seminal paper of  Fr\"{o}hlich and Spencer \cite{frsP}, the authors presented a contour argument which works even for long-range models  in one dimension, in particular, it worked for one-dimensional Dyson models with $\alpha^{*}  < \alpha  \leq  2$.  The techniques used in \cite{CFMP} rely on developing a  graphical representation  of spin configurations in terms of triangular \textit{contours}.  

It turns out that for a one-dimensional long-range model in a decaying field, depending on the relation between $\alpha$ and $\gamma$, there can be either one or two extremal Gibbs measures.
We can prove the following theorem.      
\begin{theorem}[\cite{BEEKR}]
Let  $\alpha \in (1,2]$ and $\gamma > \max \{\alpha-1,\alpha^* -1 \}$ be the exponents of the Dyson model w.r.t. an interaction $\Phi^D$ given by \eqref{potF}. Then,  there exists $\beta^D_{c,h}>0$ s.t.   for all $\beta>\beta^D_{c,h}$ we have $M_0(\beta,\alpha,\gamma) >0$, i.e. $\mu^+\neq \mu^-$.
\end{theorem}
\noindent
\textbf{Sketch of proof:}
The main idea of the proof is to extend the analysis in  \cite{CFMP,COP},  combined with \cite{LP}.  Consider a finite-volume Gibbs measure on an interval, say $\Lambda = [-N,N]$ and fix $+$-boundary conditions. Each spin configuration $\sigma$ can be uniquely mapped into a triangle configuration $\underline{T}=(T_1,...,T_n)$ where endpoints of the triangles are defined by interface points dividing plus from minus spins. Contours $\Gamma$ are collections of triangles $T_i$ such that they are in some sense \textit{well separated} from each other and subadditive, so that we  
%which will allow to 
obtain a lower bound for the energy of given triangle configuration $\underline{T}$. Phase coexistence will follow from the well-known Peierls argument for $d>1$, i.e. fromthe estimate that for for $\beta$ sufficiently large
\[
\mu_{\Lambda}^+[\sigma_o=-1]\leq
 \mu_{\Lambda}^+[\{ o\in \Gamma \}]  \leq \frac{1}{Z^+_{\Lambda}} \sum_{\Gamma \ni o} \sum_{\Gamma \text{ compatible }} e^{-\beta H(\underline{T})} < \frac{1}{2}.
\] 
The main difficulty then is to obtain a \textit{good} energetic lower bound for the Hamiltonian including the effect of the external field.   \hfill $\square$
%\end{proof}
%\begin{flushright}
%$\square$
%\end{flushright}

 Physically, an  argument explaining the statement of the theorem goes as follows: There is a competition between the effect of the pair interaction and that of the external field.  Having minus boundary conditions means that inserting a large interval $[-L, L]$ of plus spins will cost an energy of order 
$$
\sum_{|i| < L} \sum_{|j| >L} |i-j|^{-\alpha}  =   O(L^{2- \alpha}).
$$ 
However, the gain in energy due to the spins following the external magnetic field is of order 
$$
\sum_{|i| < L} |i|^{-\gamma} =O(L^{1- \gamma}).
$$
Thus, (somewhat similar to an Imry-Ma argument), we see that for $\gamma > \alpha-1$ we should expect that the field is too weak to overcome the boundary conditions and the plus and minus measures are different:  $\mu^+\neq \mu^-$. 

When the opposite case pertains, that is $\gamma < \alpha -1$,  there should be a unique Gibbs measure, with a magnetisation in the direction of the field, whatever the boundary conditions employed. We are in the process of rigorising this picture.

In fact, the analogous prediction in the 2-dimensional short-range model has been fully proved by \cite{BCCP,CV}, also giving that the critical value for $\gamma$ equals $1$,  where is possible to prove the phase transition even in the critical case,  assuming that  $h$ is small enough. Here we have the same situation, we can extend the theorem above for the case when $\gamma =  \max \{\alpha-1,\alpha^* -1 \}$ if we take $h$ small enough.

The restriction on  $\gamma$ involving $\alpha^{*}$  seems due to technical reasons, since we use arguments developed in \cite{CFMP}. Also, again based on their paper,  for technically reasons we require the nearest-neighbor term to be strong enough.\\
 However, from the physical argument sketched above, we expect that these limitations should not be required  and the argument should work, just assuming the inequality between $\gamma$ and $\alpha$ .\\

% As, similar to the short-range situation, we are  using contour methods, we first studied the  above conjecture for  the restricted values of $\alpha$ treated by \cite{CFMP} and the presence of a coexistence region of small fields (large $\gamma$ has been proved so far; we expect that the ideas of \cite{LP, BCCP,CV} can be adapted to complete the above picture \cite{BEEKR}.

% One of our results is for example:
 
% \begin{proposition}\label{Proposition 3}:
% Let $2 \geq \alpha > \alpha^{*}$ and $\gamma > \alpha -1$, where 
 %$\alpha^{*} = 3 - \frac{\ln 3}{\ln 2}$, then the Dyson model with interaction decay parameter $\alpha$ and field decay parameter $\gamma$ has a a phase transition at low temperature, implying that $ \mu^{+} \neq \mu^{-}$.
% Moreover, for $1 < \alpha < \alpha^{*}$ and $\gamma > 2- \frac{\ln 3}{\ln 2}$ the same conclusion holds. 
%\end{proposition}

%The restriction in the second part of the Proposition  seems due to technical reasons, as we expect 
%that the physical arguments given above should not be limited in their range. However, the contour arguments of \cite{CFMP}  on which we base our analysis require this limitation. 
%************************************************\\
%I add acknowledgements and try to withdraw a few ref (start with ACCN and BHR)\\
%************************************************\\
 
{\bf Acknowledgments:} 
We gratefully dedicate this paper to Chuck Newman on the occasion of  his 70th birthday. Especially AvE has over many years been inspired by and profited from his scientific contributions, his always friendly stimulating questions and discussions, and his encouraging attitude, whether it was  about Dyson models, non-Gibbsian measures,  spin glasses or any other topic from statistical mechanics.  \\
We thank J. Littin for making \cite{LP} available to us and P. Picco for very helpful discussions. \\
RB is supported by CAPES, FAPESP grant 2011/16265-8  and CNPq grants 453985/2016-5 and 312112/2015-7.
 EOE  is supported by FAPESP Grants 14/10637-9 and 15/14434-8.\\
We also thank the universities of Delft and  Groningen and STAR-NWO for making various research visits possible.

%************************************************\\
%I add acknowledgements and try to withdraw a few ref (start with ACCN and BHR)\\
%************************************************\\

\end{document}